\documentclass[aps,prl,reprint,twocolumn,superscriptaddress,floatfix,nofootinbib]{revtex4-2}
\usepackage{float}
\usepackage{microtype}
\usepackage{graphicx, amsfonts, amsthm, xr}
\usepackage{array}
\usepackage{epsfig,amsmath, amssymb, color, dsfont, upgreek, physics}
\usepackage{mathrsfs}
\usepackage{mathtools}
\usepackage{bbold}
\usepackage{comment}
\usepackage{braket}
\usepackage{lipsum}
\usepackage{soul}
\usepackage{bm}

\usepackage[bookmarks=true,colorlinks,linkcolor=OrangeRed,urlcolor=NavyBlue,citecolor=RoyalBlue]{hyperref}
\usepackage[capitalise]{cleveref}
\Crefname{figure}{Fig.}{Figs.}
\Crefname{section}{Sec.}{Secs.}
\usepackage[dvipsnames]{xcolor}
\usepackage[inline]{enumitem}
\setlist[enumerate]{label=(\roman*)}
\usepackage{cmds}
\usepackage[normalem]{ulem}
\definecolor{forestgreen}{rgb}{0.13, 0.55, 0.13}
\usepackage{orcidlink}
\newcommand{\orcidgiuseppeC}{\orcidlink{0000-0002-5749-2224}}
\newcommand{\orcidpietro}{\orcidlink{0000-0001-5279-7064}}
\newcommand{\orcidsimone}{\orcidlink{0000-0002-8882-2169}}
\newcommand{\orcidgiovanni}{\orcidlink{0000-0002-9073-8978}}
\newcommand{\orcidjad}{\orcidlink{0000-0002-0659-7990}}
\newcommand{\DFA}{\affiliation{Dipartimento di Fisica e Astronomia "G. Galilei", Università di Padova, I-35131 Padova, Italy.}}
\newcommand{\PQTC}{\affiliation{Padua Quantum Technologies Research Center, Università degli Studi di Padova}}
\newcommand{\INFNPD}{\affiliation{Istituto Nazionale di Fisica Nucleare (INFN), Sezione di Padova, I-35131 Padova, Italy.}}
\newcommand{\LMU}{\affiliation{Department of Physics and Arnold Sommerfeld Center for Theoretical Physics (ASC), Ludwig Maximilian University of Munich, 80333 Munich, Germany}}
\newcommand{\MPQ}{\affiliation{Max Planck Institute of Quantum Optics, 85748 Garching, Germany}}
\newcommand{\MCQST}{\affiliation{Munich Center for Quantum Science and Technology (MCQST), 80799 Munich, Germany}}
\graphicspath{{figures/}}
\makeatletter
\AtBeginDocument{%
  \def\selectlanguage#1{}%
  \def\otherlanguage#1{}%
}
\makeatother
\begin{document}

\title{Disorder-Free Localization and Fragmentation in a Non-Abelian Lattice Gauge Theory}
\author{Giovanni Cataldi$^{\orcidgiovanni}$}\email{giovanni.cataldi@mpq.mpg.de} \INFNPD \DFA \PQTC \MPQ \MCQST
\author{Giuseppe Calaj\'o$^{\orcidgiuseppeC}$} \DFA\INFNPD
\author{Pietro Silvi$^{\orcidpietro}$} \INFNPD \DFA \PQTC
\author{Simone Montangero$^{\orcidsimone}$} \INFNPD \DFA \PQTC
\author{Jad C.~Halimeh$^{\orcidjad}$}\email{jad.halimeh@physik.lmu.de} \MPQ \LMU \MCQST
\date{\today}

\begin{abstract}
    We investigate how isolated quantum many-body systems dynamically equilibrate under non-Abelian gauge-symmetry constraints.
    By encoding gauge superselection sectors into static $\mathrm{SU}(2)$ background charges, we map out the dynamical phase diagram of a (1+1)D $\mathrm{SU}(2)$ lattice gauge theory with dynamical matter.
    We uncover three distinct regimes: (i) an ergodic phase,
    (ii) a fragmented phase that is nonthermal but delocalized, and (iii) a disorder-free many-body localized regime.
    In the latter, a superposition of gauge superselection sectors preserves spatial matter inhomogeneities in time, as evidenced by distinct temporal scalings of entropy.
    We highlight the non-Abelian nature of these phases and argue for potential realizations on qudit processors.
\end{abstract}

\maketitle

\textbf{\textit{Introduction.---}}
Lattice gauge theories (LGTs) provide a non-perturbative formulation of continuum gauge theories \cite{Weinberg1995QuantumTheoryFields, Gattringer2009QuantumChromodynamicsLattice, Zee2003QuantumFieldTheory}, where spacetime is discretized into a grid and gauge fields are represented by link variables \cite{Rothe2012LatticeGaugeTheories}.
Originally developed to investigate quark (de)confinement in quantum chromodynamics \cite{Wilson1974ConfinementQuarks, Berges2021QCDThermalizationInitio, Gross202350YearsQuantum}, LGTs have become central to a surge of quantum simulation experiments
\cite{Martinez2016RealtimeDynamicsLattice, Klco2018QuantumclassicalComputationSchwinger, Gorg2019RealizationDensitydependentPeierls, Schweizer2019FloquetApproachZ2, Mil2020ScalableRealizationLocal, Yang2020TimedependentVariationalPrinciple, Wang2022ObservationEmergent$mathbbZ_2$, Zhou2022ThermalizationDynamicsGauge, Wang2023InterrelatedThermalizationQuantum, Zhang2025ObservationMicroscopicConfinement, Su2023ObservationManybodyScarring, Ciavarella2024QuantumSimulationSU3, Ciavarella2025StringBreakingHeavy, De2024ObservationStringbreakingDynamics, Liu2024QuantumAlgorithmsInverse, Farrell2024ScalableCircuitsPreparing, Farrell2024QuantumSimulationsHadron, Zhu2024ProbingFalseVacuum, Ciavarella2021TrailheadQuantumSimulation, Ciavarella2023QuantumSimulationLattice, Ciavarella2022PreparationSU3Lattice, Gustafson2024PrimitiveQuantumGates-1, Gustafson2024PrimitiveQuantumGates-2, Lamm2024BlockEncodingsDiscrete, Farrell2023PreparationsQuantumSimulations-1, Farrell2023PreparationsQuantumSimulations-2, Li2024SequencyHierarchyTruncation, Zemlevskiy2025ScalableQuantumSimulations, Lewis2019QubitModelU1, Atas2023SimulatingOnedimensionalQuantum, ARahman2022SelfmitigatingTrotterCircuits, Atas2021SU2HadronsQuantum, Mendicelli2023RealTimeEvolution, Kavaki2024SquarePlaquettesTriamond, Than2024PhaseDiagramQuantum, Angelides2025FirstorderPhaseTransition, Alexandrou2025RealizingStringBreaking, Cochran2025VisualizingDynamicsCharges, Gyawali2025ObservationDisorderfreeLocalization, Gonzalez-Cuadra2025ObservationStringBreaking, Crippa2026AnalysisConfinementString} in the last decade.
More recently, they have emerged as powerful tools for studying quantum many-body (QMB) dynamics \cite{Halimeh2025ColdatomQuantumSimulators}.
In particular, due to the presence of gauge symmetry \cite{Santos2010LocalizationEffectsSymmetries, Mallayya2019PrethermalizationThermalizationIsolated}, which enforces strong local constraints on dynamics,
LGTs have been successful in discovering and exploring novel types of nonthermal behavior that circumvent the Eigenstate Thermalization Hypothesis (ETH) \cite{Deutsch1991QuantumStatisticalMechanics, Srednicki1994ChaosQuantumThermalization, Rigol2006HardcoreBosonsOptical, Rigol2007RelaxationCompletelyIntegrable, Rigol2008ThermalizationItsMechanism,
    Eisert2015QuantumManybodySystems,
    DAlessio2016QuantumChaosEigenstate, Deutsch2018EigenstateThermalizationHypothesis, Mori2018ThermalizationPrethermalizationIsolated}.
In generic interacting isolated quantum systems, if ETH holds, then for initial pure states with a narrow energy distribution, local observables (or equivalently, sufficiently small subsystems) relax at long times to values predicted by the thermal ensemble fixed by the initial state energy \cite{Rigol2008ThermalizationItsMechanism, Kaufman2016QuantumThermalizationEntanglement}.
\begin{figure}[!ht]
    \includegraphics[width=1\columnwidth]{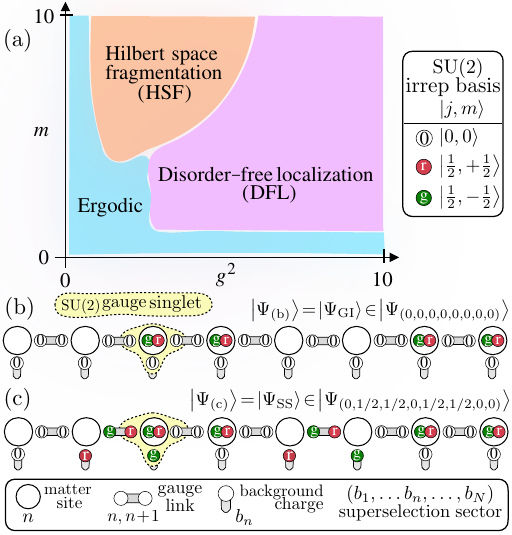}
    \caption{\textbf{Model phases and initial-states.}
        (a) Sketch of the explored dynamical phases as a function of mass and coupling strengths.
        (b)-(c) Cartoon picture of two possible charge density wave states (on $N\!=\!8$ site-chain) forming the superposition initial state in \cref{eq_bg_superposition}:
        (b) belongs to the SU(2) gauge invariant (GI) sector; (c) belongs to a superselection-sector (SS) where SU(2) Gauss law is modified with SU(2) static background charges on four lattice vertices.
        Circles, gray ovals, and bottom stubs denote matter sites, gauge links, and static background charges expressed in SU(2) irrep bases.
    }
    \label{fig_sketch}
\end{figure}
A paradigmatic ergodicity-breaking phenomenon in LGTs is QMB scarring \cite{Bernien2017ProbingManybodyDynamics, Surace2020LatticeGaugeTheories, Turner2018WeakErgodicityBreaking, Moudgalya2018ExactExcitedStates, Zhao2020QuantumManyBodyScars, Jepsen2022LonglivedPhantomHelix, Serbyn2021QuantumManybodyScars, Moudgalya2022QuantumManybodyScars, Chandran2023QuantumManyBodyScars, Bluvstein2021ControllingQuantumManybody, Bluvstein2022QuantumProcessorBased, Zhang2023ManybodyHilbertSpace, Dong2023DisordertunableEntanglementInfinite, Calajo2025QuantumManybodyScarring}, where special eigenstates, often equally spaced in energy, exhibit anomalously low entanglement entropy \cite{Moudgalya2018EntanglementExactExcited, Schecter2019WeakErgodicityBreaking, Lin2019ExactQuantumManyBody}.
Scars have been experimentally observed in LGTs \cite{Bernien2017ProbingManybodyDynamics, Su2023ObservationManybodyScarring, Desaules2024RobustFinitetemperatureManybody} and theoretically shown to persist toward the lattice quantum field theory limit \cite{Desaules2023WeakErgodicityBreaking, Desaules2023ProminentQuantumManybody}, in one and higher spatial dimensions \cite{Osborne2024QuantumManyBodyScarring, Budde2024QuantumManyBodyScars}, for Abelian \cite{Iadecola2020QuantumManybodyScar, Banerjee2021QuantumScarsZero, Halimeh2023RobustQuantumManybody, Hudomal2022DrivingQuantumManybody, Aramthottil2022ScarStatesDeconfined, Biswas2022ScarsProtectedZero, VanDamme2023AnatomyDynamicalQuantum, Daniel2023BridgingQuantumCriticality, Sau2024SublatticeScarsTwodimensional, Desaules2024MassAssistedLocalDeconfinement}, and non-Abelian gauge groups \cite{Calajo2025QuantumManybodyScarring}.

Other examples of nonthermal dynamics in LGTs are Hilbert space fragmentation (HSF) \cite{Sala2020ErgodicityBreakingArising, Khemani2020LocalizationHilbertSpace}, and disorder-free localization (DFL) \cite{Smith2017DisorderFreeLocalization, Brenes2018ManyBodyLocalizationDynamics}.
In the first case, the Hilbert space, even after resolving all global and local symmetries, shatters into dynamically disconnected Krylov subsectors --- subspaces spanned by the repeated action of the Hamiltonian on a given initial state --- whose number grows exponentially in system size \cite{Moudgalya2022QuantumManybodyScars, Moudgalya2022HilbertSpaceFragmentation}.
HSF avoids thermalization and can lead to slow or glassy dynamics confined within strict subspaces from the outset.
In the case of DFL, an initial spatial imbalance in the matter fields, which typically melts away in the absence of explicit disorder, survives in time if the system is initialized in a superposition of an extensive number of gauge superselection sectors.
DFL has been theoretically demonstrated in Abelian LGTs \cite{Smith2017AbsenceErgodicityQuenched, Smith2018DynamicalLocalization$ensuremathmathbbZ_2$, Metavitsiadis2017ThermalTransportTwodimensional, Russomanno2020HomogeneousFloquetTime, Papaefstathiou2020DisorderfreeLocalizationSimple, McClarty2020DisorderfreeLocalizationManybody, Hart2021LogarithmicEntanglementGrowth, Zhu2021SubdiffusiveDynamicsCritical, Halimeh2022StabilizingDisorderFreeLocalization, Karpov2021DisorderFreeLocalizationInteracting, Lang2022DisorderfreeLocalizationStark, Halimeh2022TemperatureInducedDisorderFreeLocalization, Chakraborty2022DisorderfreeLocalizationTransition, Sous2021PhononinducedDisorderDynamics, Halimeh2022EnhancingDisorderFreeLocalization, Homeier2023RealisticSchemeQuantum, Osborne2023DisorderFreeLocalization$2+1$D, Sala2024DisorderfreeLocalizationPurely, Halimeh2024DisorderFreeLocalizationBenchmarking, Jeyaretnam2025HilbertSpaceFragmentation}, and experimentally observed on a \texttt{Google Quantum AI} quantum processor \cite{Gyawali2025ObservationDisorderfreeLocalization}, but has not been shown in a non-Abelian LGT before now.

The type of underlying symmetries plays a central role in the equilibration of QMB systems \cite{Santos2010LocalizationEffectsSymmetries, Mallayya2019PrethermalizationThermalizationIsolated, Rigol2009BreakdownThermalizationFinite, Rigol2009QuantumQuenchesThermalization}.
Non-Abelian symmetries, such as $\mathrm{SU}(2)$ and $\mathrm{SU}(3)$, are ubiquitous in condensed matter and high-energy physics and display non-commuting conserved quantities that cannot be simultaneously resolved, and greatly enrich the underlying physics \cite{YungerHalpern2016MicrocanonicalResourcetheoreticDerivations, Popescu2020ReferenceFramesWhich, YungerHalpern2020NoncommutingConservedCharges, Manzano2022NonAbelianQuantumTransport, Patil2023AveragePurestateEntanglement, Ebner2024EigenstateThermalization$2+1$dimensional,  Yao2023SU2GaugeTheory, Bianchi2024NonAbelianSymmetryresolvedEntanglement, Ebner2024EntanglementEntropy$2+1$dimensional,
    Patil2025EigenstateThermalizationSpin$frac12$}.
Their role in QMB dynamics has generated considerable recent interest \cite{Murthy2023NonAbelianEigenstateThermalization, Majidy2023NoncommutingConservedCharges}, including a trapped-ion experiment \cite{Kranzl2023ExperimentalObservationThermalization}.

In this Letter, we investigate the role of non-Abelian symmetries in the long-time dynamics of LGTs. We observe for the first time the emergence of DFL in an $\mathrm{SU}(2)$ LGT coupled to dynamical matter, and also find evidence for HSF; see \cref{fig_sketch}.
The uncovered rich (non)thermal dynamics is revealed through several measures, such as matter imbalance, quark population, and entanglement entropy scaling, which reflect the underlying non-Abelian symmetry constraints.

\textbf{\textit{Model.---}}
As a prototypical model to observe DFL in non-Abelian LGTs, we consider a (1+1)D hardcore-gluon $\mathrm{SU}(2)$ Yang--Mills LGT coupled with flavorless dynamical matter \cite{SM, Cataldi2024Simulating2+1DSU2, Calajo2024DigitalQuantumSimulation, Calajo2025QuantumManybodyScarring}.
This model is described by the following Kogut--Susskind Hamiltonian \cite{Kogut1975HamiltonianFormulationWilsons}, on an $N$-site lattice chain with lattice spacing $\lspace$ ($\lspeed,\hbar=1$):
\begin{equation}
    \label{eq_H}
    \begin{split}
        \hat{H} & _0=\frac{1}{2\lspace}\sum_{\site}\sum_{\alpha, \beta}
        \qty[i\hpsi^{\dagger}_{\site,\alpha}\hat{U}^{\alpha\beta}_{\site,\site+1}\hpsi_{\site+1,\beta}+\hc]      \\
                & + m_0\sum_{\site}\sum_{\alpha} (-1)^{\site} \hpsi^{\dagger}_{\site,\alpha}\hpsi_{\site,\alpha}
        + \frac{\lspace g_0^2}{2}\sum_{\site}\hat{E}^2_{\site,\site+1}
        \,.
    \end{split}
\end{equation}
The model in \cref{eq_H} describes the interaction between flavorless quark matter fields of mass $m_0$, living on lattice sites $\site$,
and $\mathrm{SU}(2)$ gauge fields living on lattice links $(\site, \site\!+\!1)$, coupled with strength $g_{0}$ (see \cref{fig_sketch}(b-c)).
The quark field is represented as a staggered fermion doublet $\hpsi_{\site,\alpha}$, which satisfies the anti-commutation relations
$\acomm*{\hpsi_{\site,\alpha}}{\hpsi^{\dagger}_{\site^{\prime},\beta}}{=} \delta_{\site,\site^{\prime}} \delta_{\alpha, \beta}$ and $\acomm*{\hpsi_{\site,\alpha}}{\hpsi_{\site^{\prime}\beta}}{=}0$
\cite{Susskind1977LatticeFermions}, where $\alpha, \beta$ indices live in the fundamental $\mathrm{SU}(2)$ irreducible representation (irrep) $(\spin\!=\!1/2, m\!=\!\pm1/2)$.
The single-site matter basis is given by:
\begin{math}
    \{
    {\ket{0}},\,
    {\ket{\rla} = \hpsi^{\dagger}_{\rla}\ket{0}},\,
    {\ket{\gla} = \hpsi^{\dagger}_{\gla}\ket{0}},\,
    {\ket{2}    = \hpsi^{\dagger}_{\rla} \hpsi^{\dagger}_{\gla}\ket{0}}
    \}
\end{math},
where $\qty{\rla,\gla}$ are shorthand notations for $\qty{m\!=\!\pm1/2}$, while $\spin$ is implicit \cite{Cataldi2024Simulating2+1DSU2, Calajo2024DigitalQuantumSimulation, Calajo2025QuantumManybodyScarring} (see also \cref{fig_sketch}).

For gauge link states, we adopt the chromoelectric basis $\ket{\spin, \mL, \mR}$, where $\spin\!\in\!\mathbb{N}/2$ indicates the spin irreps and $\mR,\mL\in\{-\spin,\ldots,+\spin\}$ label the states within the spin shell $\spin$.
In this basis, the link energy density operator is diagonal and coincides with the quadratic Casimir, $\hat{E}^2\ket{\spin, \mL, \mR}=\spin(\spin\mathop+1)\ket{\spin, \mL, \mR}$ \cite{Zohar2015FormulationLatticeGauge}.
In this study, we implement the hardcore-gluon truncation and restrict $\spin \!\in\!\qty{0,1/2}$, retaining only the states that can be reached from the singlet $\ket{00}$ by applying the parallel transporter $\hat{U}^{\alpha\beta}$ at most once.
The resulting truncated basis is
\begin{math}
    \qty{\ket{00},\ket{\rla\rla},\ket{\gla\gla},\ket{\gla\rla},\ket{\rla\gla}}
\end{math}
\cite{Cataldi2024Simulating2+1DSU2, Calajo2024DigitalQuantumSimulation, Calajo2025QuantumManybodyScarring}.

\begin{figure*}
    \includegraphics[width=1\textwidth]{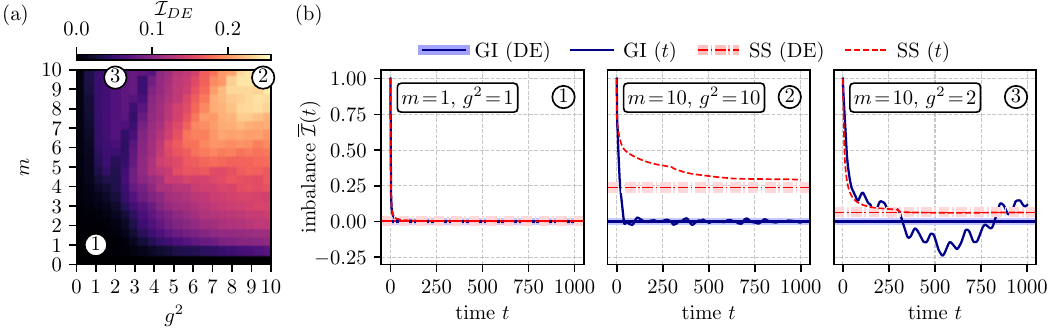}
    \caption{\textbf{Matter imbalance.} (a) DE prediction of the imbalance in \cref{eq_imbalance} as a function of mass and coupling strength in a grid with $m\in [0,10]$ and $ g^{2}\in [0,10]$.
        The three numbered white circles indicate the parameters used in panel (b) to characterize the ergodic (1), DFL (2), and HSF (3) phases in \cref{fig_sketch}.
        (b) Time evolution of the (time-averaged) imbalance obtained from the SS and GI initial states in \cref{eq_bg_superposition} (dashed red line) and \cref{eq_2siteDW} (solid blue line), respectively, for three different mass-coupling configurations.
        The averaged imbalance is always compared with the corresponding DE (horizontal lighter line) prediction.
    }
    \label{fig_imbalance}
\end{figure*}

Non-Abelian $\mathrm{SU}(2)$ gauge invariance is locally manifest at each lattice site $\site$ through generators of local rotations $\hat{G}_{\site}{=}(\hat{G}^{x}_{\site},\hat{G}^{y}_{\site},\hat{G}^{z}_{\site})$ that satisfy the following relation:
\begin{equation}
    \hat{G}_{\site}
    \ket{\Psi_{\{b^{\kappa}_{\site}\}}} =
    b_{\site}^{\kappa}
    \ket{\Psi_{\{b^{\kappa}_{\site}\}}}, \; \forall \site,\! \kappa,
    \label{eq_superselection}
\end{equation}
where $b_{\site}^{\kappa}$ is the background-charge present on the lattice site $\site$ of a QMB state $\ket{\Psi_{\{b^{\kappa}_{\site}\}}}$ within a specific \emph{gauge superselection sector} $\kappa$, \idest{}, a specific lattice configuration of background charges $\{b^{\kappa}_{\site}\}\!=\!\qty(b^{\kappa}_{1},\!\dots\!,\!b^{\kappa}_{N})$.
Similarly to the gauge links, each background charge $b_{\site}^{\kappa}$ can be expressed in an irrep basis $\ket{\spin_{b},m_{b}}$, with $\spin_{b}\!\in\!\mathbb{N}/2$ (see \cref{fig_imbalance}(b-c)).
Restricting to the first two irreps, \idest{} $\spin_{b}\!\in\!\qty{0, 1/2}$ is sufficient to detect DFL.
In numerical simulations, we locally enforce the correct background charge in \cref{eq_superselection} using a dressed site approach \cite{Cataldi2024Simulating2+1DSU2, Calajo2024DigitalQuantumSimulation, Calajo2025QuantumManybodyScarring}, which yields a defermionized \cite{Ballarin2024DigitalQuantumSimulation} qudit model with a $13$-dimensional single-site basis.
The first six states with $\ket{b_{0}}{=}\ket{\spin_{b}{=}0,m_{b}{=}0}$ correspond to the sector with no background charges (\idest{} the Gauss Law), while the remaining seven states have background charges $\qty{\ket{b_{\rla}}\!,\!\ket{b_{\gla}}}{=}\qty{\ket{\spin_{b}{=}1/2,m_{b}{=}\pm 1/2}}$.
Its derivation is detailed in the Supplemental Material (SM) \cite{SM}.
For numerical convenience, we rescale the Hamiltonian in \cref{eq_H} as $(\hat{H},m,g^2){=}4 \sqrt{2} \lspace (\hat{H}_{0},m_0,3\lspace g_0^2/16)$ \cite{Calajo2024DigitalQuantumSimulation, Calajo2025QuantumManybodyScarring} and use exact diagonalization on a chain of $N{=}8$ $\mathrm{SU}(2)$ dressed sites with periodic boundary conditions.

\textbf{\textit{Disorder-free localization.---}}
To observe DFL, we initialize the dynamics from an equal superposition of all the $\mathcal{N}_{\rm{SS}}$ gauge superselection sectors (SS) of background-charges:
\begin{equation}
    \begin{split}
        \ket{\Psi_{\rm{SS}}}
         & {=}\frac{1}{\sqrt{\mathcal{N}_{\rm{SS}}}}\sum_{\kappa=1}^{\mathcal{N}_{\rm{SS}}}
        \ket{\Psi_{\{b^{\kappa}_{\site}\}}},\;\;
        b_{\site}^{\kappa}\in \qty{0,1/2},\; \forall \site, \kappa.
    \end{split}
    \label{eq_bg_superposition}
\end{equation}
Specifically, each background-charge sector contributes with two simple translation-invariant product states where (i) matter fields are in a 2-site charge density wave configuration (staggered chain with two empty sites alternated with two doubly occupied ones) and (ii) gauge fields can be in any of the $\spin\in\qty{0,1/2}$ $\mathrm{SU}(2)$ irreps.
In this way, the initial state encodes a finite imbalance in the matter configuration under the Hamiltonian in \cref{eq_H} and evolves in a superposition of all the possible gauge field configurations.
A pictorial representation of some of these contributions is in \cref{fig_sketch}(b,c).

For direct comparison, we also consider the same matter configuration of the state in \cref{eq_bg_superposition} with no active gauge fields, which belongs to the gauge-invariant (GI) sector without background charges (see \cref{fig_sketch}(b)):
\begin{equation}
    \begin{split}
        \ket{\Psi_{\rm{GI}}} &
        \in\ket{\Psi_{\qty[0,0,\dots, 0]}}.
    \end{split}
    \label{eq_2siteDW}
\end{equation}
To detect the occurrence of DFL in the system, we introduce the following matter imbalance:
\begin{equation}
    \mathcal{I}(t)=\frac{1}{N}\sum_{n=1}^{N}w_n\rho_{\site}(t)\,,
    \label{eq_imbalance}
\end{equation}
where $0\leq \rho_{\site}(t)\leq 2$ is the single-site fermion density:
\begin{equation}
    \rho_{\site}(t)\equiv\sum_{\alpha}\mel{\Psi(t)}{\hat{\psi}_{\site,\alpha}^{\dagger}\hat{\psi}_{\site,\alpha}}{\Psi(t)},
\end{equation}
and the coefficients $w_n\!=\!\rho_{\site}(0)\!-\!1$ define the weights associated with the considered initial matter configuration.
The matter imbalance in \cref{eq_imbalance} quantifies how much the system departs from or retains the matter configuration of the initial state.
To compute the long-time average of this and any other local observable $\hat{O}(t)$, we use the diagonal ensemble (DE) \cite{Rigol2008ThermalizationItsMechanism, SM}, which, for the initial state expressed in the Hamiltonian eigenbasis, $\ket{\Psi(0)}\!=\!\sum_{s}C_{s}\ket{\Phi_{s}}$, the long-time average is  $O_{\mathrm{DE}}\!=\!\sum_{s}\abs{C_{s}}^2 \braket{\Phi_s|\hat O|\Phi_s}$.
In \cref{fig_imbalance}(a), we plot the DE imbalance $\mathcal{I}_{\mathrm{DE}}$ for the SS state of \cref{eq_bg_superposition} as a function of mass and gauge coupling strength.
The plot shows that for large coupling and moderately large masses, $(g\!\gg\!1, m\!\gtrsim\!1)$, the long-time average of the imbalance $\mathcal{I}_{DE}$ acquires finite values, signaling long-time localization of the initial configuration.

The dynamics of the imbalance approaching the long-time average are shown in \cref{fig_imbalance}(b).
There, to mitigate the effect of finite-size oscillations, we consider the time-averaged imbalance $\overline{\mathcal{I}}(t)=\frac{1}{\Delta t}\int^{t+\Delta t/2}_{t-\Delta t/2}ds\,\mathcal{I}(s)$, for an intermediate time window $\Delta t=300$.
We focus on three parameter points (1-3 circles in \cref{fig_imbalance}(a)) representing the ergodic (1), DFL (2), and HSF (3) regions sketched in \cref{fig_sketch}(a), and compare dynamics from the GI and SS initial states.
For the GI initial state, the imbalance $\overline{\mathcal I}(t)$ does not remain finite at long times in any of the three regimes; in the HSF region, it relaxes more slowly than in the ergodic case.
In contrast, for the SS initial state, the ergodic region shows decay of the imbalance, whereas in the HSF and DFL regions $\overline{\mathcal I}(t)$ approaches a nonzero DE value, indicating persistent memory of the initial matter pattern.
These results suggest that the model may exhibit DFL in specific parameter regimes, as we show in more detail in the following analysis.
\begin{figure*}
    \includegraphics[width=1\textwidth]{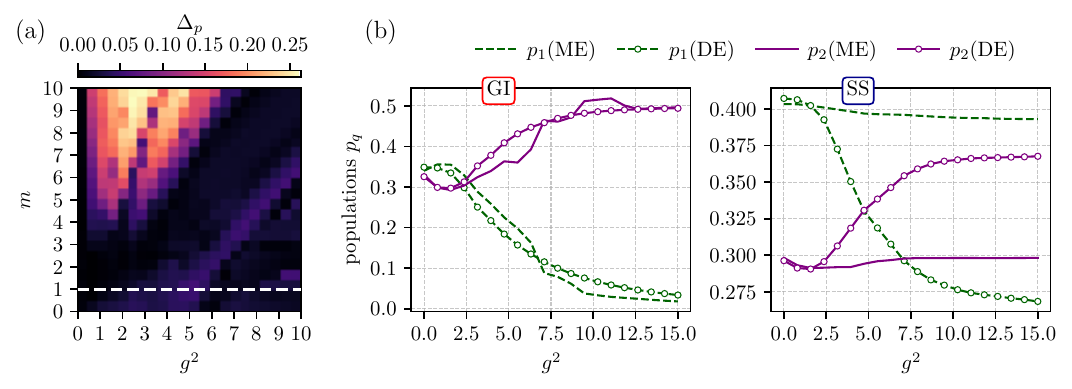}
    \caption{\textbf{Long-time dynamics of quark occupancy.} (a) Mass-gauge coupling grid of the deviation $\Delta p$ of the DE and ME predictions, as defined in \cref{eq_delta_p}, of the single $p_{1}$ and double $p_{2}$ quark populations for the case of the gauge-invariant (GI) initial state.
        (b) Average quark populations obtained from ME and DE predictions for the GI and SS states at fixed mass $m=1$ and gauge coupling strength $g^{2} \in [0, 15]$ (along the white dashed line in panel (a)).}
    \label{fig_population}
\end{figure*}

\textbf{\textit{Particle creation.---}}
To confirm that the observed localization arises from the superposition of different superselection sectors, we must first ensure that the GI system (without background charges) exhibits ergodic behavior.
We define the single-site occupancy operator
\begin{equation}
    \hat{Q}_{\site} = (-1)^{\site}\sum_{\alpha}
    \qty[\hpsi^{\dagger}_{\site,\alpha}\hpsi_{\site,\alpha}-\frac{1-(-1)^{\site}}{2}]\,,
\end{equation}
which, in the staggered fermion description \cite{Susskind1977LatticeFermions, Calajo2024DigitalQuantumSimulation, Calajo2025QuantumManybodyScarring} counts the number of quarks (on even sites) or antiquarks (on odd sites).
Then, we compute the average quark population $p_{q}(t)$ of each $\hat{Q}_n$ eigenvalue (single-site quark occupancy, $q\in\{0,1,2\}$), projecting on the corresponding eigensubspace of $\hat Q_n$ and averaging over all sites $\site$:
\begin{equation}
    \label{eq_population}
    p_{q}(t) = \frac{1}{N} \sum_{\site}\ev{\delta_{\hat{Q}_{\site}, q}}{\Psi(t)}\,,
\end{equation}
where the Kronecker delta $\delta_{\hat{Q}_{\site}, q}$ selects a specific occupancy.
In \cref{fig_population}(a), we quantify the deviation between the DE and the microcanonical ensemble (ME) predictions for the quark populations of the GI state \cref{eq_2siteDW}.
The ME value, which reflects the thermal prediction, is obtained by uniformly averaging $\mel{\Phi_s}{\delta_{\hat{Q}_{\site},q}}{\Phi_s}$ over eigenstates $\ket{\Phi_s}$ in a narrow energy window around the initial state energy (see SM).
In particular, we focus on the average deviation of the single $p_1$ and double $p_2$ quark occupancies
\begin{equation}
    \Delta_{p}\equiv \frac{\abs{p_{1}^\text{DE}-p_{1}^\text{ME}}+\abs{p_{2}^\text{DE}-p_{2}^\text{ME}}}{2},
    \label{eq_delta_p}
\end{equation}
as a function of the mass and gauge coupling.
In the regime of $m\gg 1, g^{2}\gtrsim 1$, there is a significant region where the ME and DE predictions deviate, indicating the absence of thermalization in the GI sector.
In SM \cite{SM}, we analyze this regime in detail and attribute the observed nonergodic behavior to HSF \cite{Sala2020ErgodicityBreakingArising, Moudgalya2022QuantumManybodyScars}, which is known to be prevalent in LGTs \cite{Jeyaretnam2025HilbertSpaceFragmentation, Ciavarella2025GenericHilbertSpace}.
Since this region is inherently nonthermal in the GI sector due to the Hamiltonian itself being nonergodic in this regime, it should be excluded from the preliminary DFL phase diagram shown in \cref{fig_imbalance}(a), since in our case, DFL is expected to possibly occur only in regimes where the Hamiltonian is ergodic at least in the GI sector.
Doing so, we identify the different dynamical phases explored by the initial state \cref{eq_bg_superposition} as the ones depicted in \cref{fig_sketch}(c).

To gain deeper insight into the nature of excitations driving the DFL phase observed in \cref{fig_imbalance}, we compute the average quark occupancy on the lattice, distinguishing between the single $p_1$ and double $p_2$ occupancies defined in \cref{eq_population}.
In \cref{fig_population}(b), we compare the ME and DE predictions for the GI and SS cases by fixing the mass at a value far from the fragmented phase and varying the gauge coupling.
In agreement with our previous analysis, in the GI sector the DE and ME predictions closely follow each other, confirming the thermal behavior of the initial state in \cref{eq_2siteDW}.
Conversely, when the system is initialized in a superposition of SS in this regime, the ME and DE predictions diverge strongly due to the emergence of the previously discussed DFL phase.
As the gauge coupling increases, the dominant contribution of the dynamics comes from the double-occupancy population $p_{2}$, which corresponds to the formation of baryon-like (quark pair) and antibaryon-like (antiquark pair) excitations.
This starkly contrasts the ME prediction, which instead anticipates a greater production of meson-like excitations (adjacent quark-antiquark pairs with an excited shared gauge link) as indicated by the single-occupancy population $p_{1}$.
The crucial role of baryonic excitations is a distinctive feature of the non-Abelian $\mathrm{SU}(2)$ model, setting this phenomenology apart from its Abelian counterpart \cite{Brenes2018ManyBodyLocalizationDynamics}.

\begin{figure}
    \includegraphics[width=1\columnwidth]{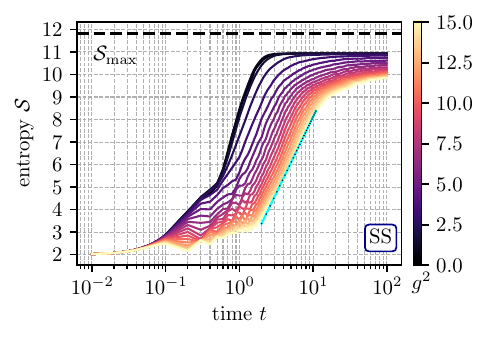}
    \caption{\textbf{Entanglement scaling.}
    Half-chain bipartite entanglement entropy in base two as a function of time (in logarithmic scale) for fixed mass $m\!=\!1$, and a range of gauge couplings $g^{2}\!\in\![0, 15]$ in the superposition of superselection sectors (SS) case.
    The cyan dashed line is the linear fit of the $g^{2}\!=\!15$ curve and corresponds to $\mathcal{S}\!=\!2.94(2)\!\cdot\!\log(t)\!+\!1.33(3)$.
    The black dashed line represents the maximal bipartite entanglement entropy value $\mathcal{S}_{\rm max}\!$ corresponding to an equal distribution over all accessible system configurations.
    Notice that the SS initial state given in \cref{eq_bg_superposition} under periodic boundary conditions is not a product state, as in the Abelian case \cite{Brenes2018ManyBodyLocalizationDynamics}, rather, an exact matrix product state with bond dimension $\chi\!=\!4$ \cite{SM}, thus the entanglement entropy at the initial time starts from $\mathcal{S}\!=\!\log\chi$.}
    \label{fig_entropy}
\end{figure}

\textbf{\textit{Entanglement Entropy.---}}
A hallmark feature of DFL is the slow spreading of quantum correlations across the system, reminiscent of that in a many-body localized (MBL) phase \cite{DeChiara2006EntanglementEntropyDynamics, Brenes2018ManyBodyLocalizationDynamics, Znidaric2008ManybodyLocalizationHeisenberg, Pal2010ManybodyLocalizationPhase, Bardarson2012UnboundedGrowthEntanglement, Huse2014PhenomenologyFullyManybodylocalized}.
Indeed, while in ergodic systems governed by the ETH, quantum correlations spread linearly with time, in MBL phases, entanglement entropy grows logarithmically with time.
This phenomenon can be quantified by measuring the bipartite von Neumann entanglement entropy, defined as $S\!=\!-{\Tr}[\hat{\varrho}_A \log \hat{\varrho}_A]$, where $A$ and $B$ denote the two halves of the chain described by the density operator $\hat{\varrho}_{AB}$, and $\hat\varrho_A\!=\!\Tr_{B}[\hat{\varrho}_{AB}]$ is the reduced density operator of subsystem $A$.
In \cref{fig_entropy}, we show the entanglement entropy scaling for the SS initial state across a range of gauge couplings $g^{2}\!\in\![0,15]$ at a fixed mass $m\!=\!1$.
Aside from late-time saturation effects due to the finite system size, the SS case exhibits a clear linear entanglement growth in $\log(t)$ across almost all the coupling values, compatible with an MBL phase and similar to what was observed in the Abelian scenario \cite{Brenes2018ManyBodyLocalizationDynamics}.
In detail, stronger gauge couplings limit the spread of entanglement, leading to lower saturation values.
Indeed, as $g^{2}$ increases, gauge-link excitations are strongly penalized, and hopping processes are thereby suppressed.
This constrained dynamics affects the spread of quantum correlations in each superselection sector, including the GI case, where the entanglement entropy scales linearly in time (see SM \cite{SM}), which is compatible with ergodicity.

\textbf{\textit{Summary and outlook.---}}
We have investigated the dynamical phases of a (1+1)D truncated $\mathrm{SU}(2)$ LGT with dynamical matter.
To explore the dynamics across different gauge superselection sectors, we developed a formalism based on static $\mathrm{SU}(2)$ background charges, enabling the preparation of superpositions across distinct sectors.
Within this framework, we identified three different regimes: in addition to a thermal ergodic phase, we found a nonthermal yet delocalized fragmented regime and a DFL phase that emerges when superpositions of gauge superselection sectors are considered.
In the latter, an initial imbalance in the matter distribution persists over time, and the entanglement entropy grows logarithmically, reminiscent of MBL.
The dynamics in these regimes are strongly shaped by non-Abelian excitations such as baryons.
In the DFL regime, characterized by large gauge coupling and mass, the gauge-field and background-charge truncations are well controlled.
We expect the qualitative features of the dynamics to remain robust under larger truncation cutoffs, as supported by recent results on QMB scars in the same model~\cite{Calajo2025QuantumManybodyScarring}.

Several immediate directions follow from our work, such as the stability of the DFL phase in (2+1)D and under different non-Abelian gauge groups.
Another interesting avenue is to analyze the robustness of DFL to coherent error terms that explicitly break the non-Abelian gauge symmetry, which are commonly encountered in quantum simulators.
The effect of decoherence on this phase is also a critical direction worth investigating.

Finally, we point out that a qudit formulation with a 13-dimensional local (single-site) Hilbert space faithfully captures the DFL phenomenology described in this work.
This makes the model particularly suitable for digital quantum simulation on qudit-based quantum processors, as recently proposed for the same model in the absence of background charges~\cite{Calajo2024DigitalQuantumSimulation}.
Platforms supporting such large local dimensions are already experimentally accessible, as demonstrated by recent implementations using 13-level trapped-ion qudits based on $^{137}\textrm{Ba}^+$\cite{Low2023ControlReadout13level}, as well as other qudit architectures employing metastable states of ion isotopes \cite{Benhelm2008ExperimentalQuantuminformationProcessing} and circular Rydberg atoms~\cite{Kruckenhauser2022HighdimensionalSO4symmetricRydberg, Cohen2021QuantumComputingCircular}.

\textbf{\textit{Acknowledgments.---}}
The authors thank Giuseppe Magnifico, Marco Rigobello, and Darvin Wanisch for fruitful discussions.
Authors acknowledge financial support from: the European Union via QuantERA2017 project QuantHEP, via QuantERA2021 project T-NiSQ, via the Quantum Technology Flagship project PASQuanS2, and the NextGenerationEU project CN00000013 - Italian Research Center on HPC, Big Data, and Quantum Computing (ICSC);
the Italian Ministry of University and Research (MUR) via PRIN2022-PNRR project TANQU, via Progetti Dipartimenti di Eccellenza project Frontiere Quantistiche (FQ), and project Quantum Sensing and Modelling for One-Health (QuaSiModO);
the Max Planck Society, the Deutsche Forschungsgemeinschaft (DFG, German Research Foundation) under Germany's Excellence Strategy – EXC-2111 – 390814868, and the European Research Council (ERC) under the European Union's Horizon Europe research and innovation program (Grant Agreement No.~101165667)—ERC Starting Grant QuSiGauge; the WCRI-Quantum Computing and Simulation Center (QCSC) of Padova University;
Regione Veneto via program PR Veneto FESR 2021-2027 project CONVECS.
This work is part of the Quantum Computing for High-Energy Physics (QC4HEP) working group.
The authors also acknowledge the computational resources of Cloud Veneto.

\textbf{\textit{Data availability.---}}
Numerical simulations have been performed with the \textrm{ed-lgt} Library for Exact Diagonalization of Lattice Gauge Theories \cite{Cataldi2026EdlgtExactDiagonalization}.
Simulation data are available on Zenodo \cite{Cataldi2026DataCodeDisorderFree}.

\bibliography{bibliography}

\clearpage
\onecolumngrid
\begin{center}
    \textbf{\large Supplemental Online Material for \\``Disorder-Free Localization and Fragmentation in a Non-Abelian Lattice Gauge Theory'' }\\[5pt]
    \vspace{0.1cm}
    \begin{quote}
        {\small In this Supplemental Material, we detail the model derivation and the numerical techniques employed in this study.}\\[10pt]
    \end{quote}
\end{center}
\setcounter{equation}{0}
\setcounter{figure}{0}
\setcounter{table}{0}
\setcounter{page}{1}
\setcounter{section}{1}
\makeatletter
\renewcommand{\theequation}{S\arabic{equation}}
\renewcommand{\thefigure}{S\arabic{figure}}
\renewcommand{\thesection}{S\Roman{section}}
\renewcommand{\thepage}{\arabic{page}}
\renewcommand{\thetable}{S\arabic{table}}
\vspace{0cm}
\twocolumngrid
\normalsize

\section{Hardcore-gluon qudit model with Background charges}
To map the Kogut--Susskind $\mathrm{SU}(2)$ lattice Yang--Mills Hamiltonian of Eq.(1) to the qudit model used in numerical simulations, we:
\begin{enumerate*}
    \item perform a hardcore-gluon truncation of the $\mathrm{SU}(2)$ gauge fields \cite{Cataldi2024Simulating2+1DSU2};
    \item build the gauge-singlet local dressed basis accounting for the presence of eventual background charges; and
    \item gauge-defermionize.
\end{enumerate*}
The procedure is conveniently formulated by decomposing matter and gauging each Hilbert space in $\mathrm{SU}(2)$ irreducible representations (irreps).
This preliminary step already appears in the bases introduced in the main text: for gauge links, we adopt the irrep basis $\ket{\spin, \mL, \mR}$; for matter sites, the Fock basis
\begin{math}
    \{\ket{q}\}
    =
    \{
    {\ket{0}},\,
    {\ket{\rla},\ket{\gla} = \hpsi^{\dagger}_{\rla,\gla}\ket{0}},\,
    {\ket{2}    = \hpsi^{\dagger}_{\rla} \hpsi^{\dagger}_{\gla}\ket{0}}
    \}
\end{math}.
Matter basis states are associated with the following spin labels $(j,m)$:
\begin{equation}
    \ket{0}               \leftrightarrow (0,0)
    \,\quad
    \ket{\rla},\ket{\gla} \leftrightarrow (1/2,\pm1/2)
    \,\quad
    \ket{2}               \leftrightarrow (0,0)
    \,.
\end{equation}
Irrep decomposition specifies how local gauge rotations act on a site $\site$ and its neighboring links.
The generators of the infinitesimal rotations read $\forall \nu\in \{x,y,z\}$:
\begin{equation}
    \hat{G}^{\nu}_{\site}=\hat{R}_{\site-1,\site}^{\nu} + \hat{Q}_{\site}^{\nu}+\hat{L}_{\site,\site+1}^{\nu}\,,
\end{equation}
where
\begin{math}
    \hat{Q}^{\nu}_{\site}{=}
    \sum_{\alpha\beta}
    \hpsi^{\dagger}_{\site,\alpha}
    S^{(1/2)\nu}_{\alpha\beta}
    \hpsi_{\site,\beta}
\end{math}
rotates the quark field at $\site$, while $\hat{R}_{\site-1,\site}^{\nu}$ and $\hat{L}^{\nu}_{\site,n+1}$ account for the transformation of the gauge links at its left and right \cite{Zohar2015QuantumSimulationsLattice}:
\begin{align}
    \label{eq_LR_rishon_rotations}
    \mel{\spin^{\prime} \mL^{\prime} \mR^{\prime}}{\hat{L}^{\nu}}{ j \mL \mR } & =\delta_{\spin,\spin^{\prime}}S^{(j)\nu}_{\mL^{\prime},\mL} \delta_{\mR^{\prime},\mR}\,, \\
    \mel{\spin^{\prime} \mL^{\prime} \mR^{\prime} }{\hat{R}^{\nu}}{j \mL \mR } & =\delta_{\spin,\spin^{\prime}}\delta_{\mL^{\prime},\mL} S^{(j)\nu}_{\mR^{\prime},\mR}\,;
\end{align}
where $S^{(j)\nu}$ are the spin-$\spin$ $\mathfrak{su}(2)$ matrices.
From these operators, we can build the chromoelectric energy operator
\begin{math}
    \hat{E}^2 =\sum_{\nu}(\hat{R}^{\nu})^2 =\sum_{\nu}(\hat{L}^{\nu})^2
\end{math},
\idest{}, quadratic Casimir \cite{Zohar2015QuantumSimulationsLattice}:
\begin{align}
    \hat{E}^2\ket{j \mL \mR } & =j(j+1)\ket{j \mL \mR }\,.
\end{align}
Correspondingly, the action of the parallel transporter is given in terms of Clebsch-Gordan coefficients \cite{Zohar2015FormulationLatticeGauge}:
\begin{equation}
    \label{eq_parallel_transporter}
    \mel{\spin^{\prime}\mL^{\prime}\mR^{\prime}}{\hat{U}^{\alpha\beta}}{\spin\mL \mR}
    =\sqrt{\frac{2\spin+1}{2\spin ^{\mathrlap{\prime}}+1}}\:
    \overline{C^{\spin,\mL}_{\jonehalf,\alpha;\,\spin^{\prime},\mL^{\prime}}}
    C^{\spin^{\prime},\mR^{\prime}}_{\jonehalf,\beta;\,\spin,\mR}.
\end{equation}
Then, at each site $\site$, the $\mathrm{SU}(2)$ Gauss law reads:
\begin{equation}
    \hat{G}_{\site}^{\nu}
    \ket{\Psi_{\qty{b^{\kappa}_{n}}}} =
    b_{\site}^{\nu,\kappa}
    \ket{\Psi_{\qty{\vb{b}^{\kappa}_{\site}}}}, \; \forall \site, \nu,\kappa,
\end{equation}
where $b_{\site}^{\nu,\kappa}$ are the static background-charges present on the lattice site $\site$ of a QMB state with a specific $\kappa$ \emph{super-selection sector}, \idest{}, a specific lattice configuration of background charges $\qty[b^{\kappa}_{1}\dots b^{\kappa}_{N}]$ (where for simplicity we omit the spin-component index $\nu$).

\subsection{Hardcore-gluon and background charge truncations}
In principle, the gauge field and the static background charges could occupy arbitrarily high spin shells.
To deal with a finite gauge-link Hilbert space, we perform the \emph{hardcore-gluon} approximation, which bounds $\spin\in\{0,1/2\}$ and yields an energy cutoff $\norm*{\hat{E}^2} \leq 3/4$ on the Casimir spectrum ($\norm{\:\cdot\:}$ denotes the matrix norm) \cite{Cataldi2024Simulating2+1DSU2, Calajo2024DigitalQuantumSimulation, Calajo2025QuantumManybodyScarring}.
A similar truncation is performed on the irreps of the static background charges $(\spin_{b},m_{b})$ where $\spin_{b}\in\{0,1/2\}$.

Recovering the untruncated gauge group is \emph{not} essential for the present investigation.
The hardcore-gluon approximation faithfully captures the relevant low-energy physics in the strong-coupling regime $g_{0} \gg 1$, where the chromoelectric energy dominates the Hamiltonian.
Remarkably, this is precisely the regime where the distinctive DFL behavior emerges.
Similarly, restricting the background charges to the lowest two spin irreps is already sufficient to resolve the onset and structure of DFL.

\subsection{Dressed-site basis with background charges}
Regardless of the gauge-field and the background-charge truncations, we can always observe that $\hat{L}^{\nu}$ and $\hat{R}^{\nu}$ from \cref{eq_LR_rishon_rotations} act nontrivially only on the $\mL$ and $\mR$ index, respectively.
We can then factorize each gauge link into a pair of new rishon degrees of freedom that live at its edges \cite{Silvi2014LatticeGaugeTensor}.
The basis states of each rishon mode are labeled as
\begin{math}
    \ket{0}\!=\!\ket{j\,{=}\,0,m\,{=}\,0}
\end{math}
and
\begin{math}
    \ket{\rla},
    \ket{\gla}\!=\!\ket{j\,{=}\,\frac{1}{2},m\,{=}\,\pm\frac{1}{2}}
\end{math}.
Combining each matter site with its two adjacent (left and right) gauge-rishons and the attached background charge, we forge a composite dressed-site with a 13-dimensional single dressed-site Hilbert space
\begin{math}
    \{
    \ket{\ell}{=}
    \ket{b^{\ell}_{\site}}\ket{\mR^{\ell}({n-1,n}),\,q^{\ell}(\site),\,\mL^{\ell}({\site,\site+1})}
    \}_{\ell=1}^{13}
\end{math}
where Gauss Law becomes an internal constraint \cite{Cataldi2024Simulating2+1DSU2}.
The first 6 basis states belong to the gauge-invariant sector with zero background-charge $b_{0}{=}\ket{\spin_{b}=0,m_{b}=0}$ \cite{Calajo2024DigitalQuantumSimulation, Calajo2025QuantumManybodyScarring}:
\begin{equation}
    \label{eq_extended_basis1}
    \begin{aligned}
        \ket{1} & \!=\!\ket{b_{0}}\!\ket{0,0,0},                                              &
        \ket{2} & \!=\!\ket{b_{0}}\!\frac{\ket{\rla,0,\gla}\!-\!\ket{\gla,0,\rla}}{\sqrt{2}},   \\
        \ket{3} & \!=\!\ket{b_{0}}\!\frac{\ket{0,\rla,\gla}\!-\!\ket{0,\gla,\rla}}{\sqrt{2}}, &
        \ket{4} & \!=\!\ket{b_{0}}\!\frac{\ket{\gla,\rla,0}\!-\!\ket{\rla,\gla,0}}{\sqrt{2}},   \\
        \ket{5} & \!=\!\ket{b_{0}}\!\ket{0,2,0},                                              &
        \ket{6} & \!=\!\ket{b_{0}}\!\frac{\ket{\rla,2,\gla}\!-\!\ket{\gla,2,\rla}}{\sqrt{2}},
    \end{aligned}
\end{equation}
while the remaining 7 states belong to the sector with finite background-charge $\qty{b_{\rla},b_{\gla}}{=}\ket{\spin_{b}=1/2,m_{b}=\pm1/2}$:
\begin{equation}
    \label{eq_extended_basis2}
    \begin{aligned}
        \ket{7}  & =\frac{\ket{b_{\rla}}\ket{0,0,\gla}\!-\!\ket{b_{\gla}}\ket{0,0,\rla}}{\sqrt{2}},  \\
        \ket{8}  & =\frac{\ket{b_{\rla}}\ket{\gla,0,0}\!-\!\ket{b_{\gla}}\ket{\rla,0,0}}{\sqrt{2}},  \\
        \ket{9}  & =\frac{\ket{b_{\rla}}\ket{0,\gla,0}\!-\!\ket{b_{\gla}}\ket{0,\rla,0}}{\sqrt{2}},  \\
        \ket{10} & =\frac{
            \ket{b_{\rla}}\qty[\ket{\gla,\rla,\gla}
                {-}\ket{\gla,\gla,\rla}]
            {-}\ket{b_{\gla}}\qty[\ket{\rla,\rla,\gla}
                {-}\ket{\rla,\gla,\rla}]
        }{2},                                                                                        \\
        \ket{11} & =\frac{
            \ket{b_{\rla}}\ket{\rla,\gla,\gla}
        +\ket{b_{\gla}}\ket{\gla,\rla,\rla}}{\sqrt{3}}                                               \\
                 & -\frac{
            \ket{b_{\rla}}\qty[\ket{\gla,\rla,\gla}{+}\ket{\gla,\gla,\rla}]{+}
            \ket{b_{\gla}}\qty[\ket{\rla,\rla,\gla}{+}\ket{\rla,\gla,\rla}]
        }{2\sqrt{3}},                                                                                \\
        \ket{12} & =\frac{\ket{b_{\rla}}\ket{0,2,\gla}\!-\!\ket{b_{\gla}}\ket{0,2,\rla}}{\sqrt{2}},  \\
        \ket{13} & =\frac{\ket{b_{\rla}}\ket{\gla,2,0}\!-\!\ket{b_{\gla}}\ket{\rla,2,0}}{\sqrt{2}} .
    \end{aligned}
\end{equation}
Notice that, since $\forall \site$, $[\hat{G}_{\site}-b_{\site},\ham]=0$, the two background-charge sectors of \cref{eq_extended_basis1,eq_extended_basis2} are completely decoupled and do not interact with each other.
Therefore, any superselection sector $\kappa$, \idest{} lattice configuration of local (single-site) background charges $\qty{b^{\kappa}_{n}}=\qty[b^{\kappa}_{1}\dots b^{\kappa}_{N}]$ is independent of the others $\{b^{\kappa^{\prime}}_{n}\}$ and can be simulated independently.

\subsection{Gauge defermionization}
The dressed-site formalism introduced previously is equivalent to the original LGT description, provided that it recovers the original physical space.
Namely, it requires that the left and right rishons on each link must be in the same spin shell $\spin$ by restricting to the even parity sector of a $\mathbb{Z}_2$ symmetry, whose action reads:
\begin{equation}
    \label{eq_link_symmetry}
    \hat{P}= +\op{0} - (\op{\rla} + \op{\gla})
    \,.
\end{equation}
Correspondingly, the parallel transporter in \cref{eq_parallel_transporter} can be written as:
\begin{equation}
    \label{eq_parallel_transporter_rishons}
    \hat{U}^{\alpha\beta}_{\site,n+1} \to \frac{1}{\sqrt{2}}
    \rishon^{\mathrm{(L)} \alpha}_{\site,n+1}
    (\rishon^{\mathrm{(R)} \beta}_{\site,n+1})^\dagger
    \,,
\end{equation}
where the rishon modes read:
\begin{equation}
    \rishon^{\rla} = \ketbra{0}{\rla} + \ketbra{\gla}{0}
    \,,\quad
    \rishon^{\gla} = \ketbra{0}{\gla} - \ketbra{\rla}{0}
    \,.
\end{equation}
Observe that the right-hand side of \cref{eq_parallel_transporter_rishons} preserves the link parity in \cref{eq_link_symmetry}, as desired.
Conversely, a single rishon operator always inverts it:
\begin{math}
    \acomm*{\hat{P}}{\rishon^{\alpha}} = 0
\end{math}.
The existence of a $\mathbb{Z}_2$-symmetry in the rishon spaces allows to gauge-defermionize the model \cite{Cataldi2024Simulating2+1DSU2, Ballarin2024DigitalQuantumSimulation}:
we take $\rishon^{\alpha}$ to be fermionic and rewrite the Hamiltonian in Eq.(1) in terms of the following bosonic operators:
\begin{equation}
    \begin{aligned}
        \hat{Q}_{\site}^{\mathrm{(L,R)}} & = \sum_{\alpha} (\rishon^{\mathrm{(L,R)} \alpha}_{\site,\site+1})^\dagger \hpsi_{\site,\alpha}
        \,,                                                                                                                               \\
        \hat{M}_{\site}                  & =\sum_{\alpha} \hpsi_{\site,\alpha}^\dagger\hpsi_{\site,\alpha},                               \\
        \hat{C}_{\site}                  & =\frac{1}{2}\sum_{\nu}\qty[\hat{R}^{\nu,2}_{\site-1, \site}+\hat{L}^{\nu,2}_{\site, \site+1}]
        \,.
    \end{aligned}
\end{equation}
For potential quantum simulation implementations, it is convenient to rewrite these solely in terms of Hermitian operators, introducing \cite{Calajo2024DigitalQuantumSimulation}:
\begin{equation}
    \begin{aligned}
        \hat{A}^{(1)} & \mathbin=\hat{Q}^{\mathrm{(L)}} \mathop+ \hat{Q}^{\mathrm{(L)}\dagger}\,,                 &
        \hat{B}^{(1)} & \mathbin=\hat{Q}^{\mathrm{(R)}} \mathop+ \hat{Q}^{\mathrm{(R)}\dagger}\,,                   \\
        \hat{A}^{(2)} & \mathbin= i \big[ \hat{Q}^{\mathrm{(L)}} \mathop- \hat{Q}^{\mathrm{(L)}\dagger} \big] \,, &
        \hat{B}^{(2)} & \mathbin= i \big[ \hat{Q}^{\mathrm{(R)}} \mathop- \hat{Q}^{\mathrm{(R)}\dagger} \big] \,.
    \end{aligned}
\end{equation}
Finally, we map Eq.(1) to the following 13-dimensional (6 with no background charge + 7 with finite background charge) qudit Hamiltonian \cite{Calajo2024DigitalQuantumSimulation}:
\begin{equation}
    \label{eq_Heff}
    \ham
    = \sum_{\site,p} \hat{A}^{(p)}_{\site} \hat{B}^{(p)}_{\site+1}
    + \mass \sum_{\site}(-1)^{\site} \hat{M}_{\site}
    + g^2 \sum_{\site} \hat{C}_{\site}\,,
\end{equation}
where we rescaled energy to absorb the hopping pre-factor, $\ham=4\sqrt{2}\lspace\ham_{0}$, and defined
the dimensionless couplings $m=4\sqrt{2}\lspace \mass[0]$, and $\coupling^{2}=\frac{3\sqrt{2}}{4}\lspace\coupling[0]^2$.

\subsection{Particle populations}
Within this dressed-site basis, we can still easily measure various particle-density operators, which yield information about the physical processes occurring during the dynamics.
In particular, we can define:
\begin{subequations}
    \begin{align}
        \hat{\rho}_{\site}^{[2]} &
        =\hpsi_{\site,\rla}^{\dagger}
        \hpsi_{\site,\rla}
        \hpsi_{\site,\gla}^{\dagger}
        \hpsi_{\site,\gla}                         \\
        \hat{\rho}_{\site}^{[1]} &
        =\hat{M}_{\site}-2\hat{\rho}_{\site}^{[2]} \\
        \hat{\rho}_{\site}^{[0]} &
        =1-\hat{\rho}^{[1]}_{\site}-\hat{\rho}_{\site}^{[2]},
    \end{align}
\end{subequations}
where $\hat{\rho}_{\site}^{[0]}$, $\hat{\rho}_{\site}^{[1]}$, and $\hat{\rho}_{\site}^{[2]}$ measure the single-site zero, one, and double occupancy respectively.
Correspondingly, within the adopted staggered fermion solution \cite{Susskind1977LatticeFermions}, these single-site observables allow us to compute true-particle quantities like the average (anti-)baryon, meson, and vacuum densities defined in Eq.(8) of the main text, which explicitly read:
\begin{subequations}
    \begin{align}
        \hat{p}_2 & =\frac{1}{\Nsites}\sum_{\site}\qty[
            \frac{1+(-1)^{\site}}{2}\hat{\rho}_{\site}^{[2]}+
        \frac{1-(-1)^{\site}}{2}\hat{\rho}_{\site}^{[0]}],                  \\
        \hat{p}_1 & =\frac{1}{\Nsites}\sum_{\site}\hat{\rho}_{\site}^{[1]}, \\
        \hat{p}_0 & =\frac{1}{\Nsites}\sum_{\site}\qty[
            \frac{1-(-1)^{\site}}{2}\hat{\rho}_{\site}^{[2]}+
            \frac{1+(-1)^{\site}}{2}\hat{\rho}_{\site}^{[0]}].
    \end{align}
    \label{eq_particle_densities}
\end{subequations}

\section{Initial state candidates for dynamics}\label{sec_intial_states}
To observe DFL, we start the dynamics with an equal superposition of all the possible configurations of single-site background charges, \idest{}, a superposition of all the $\mathcal{N}_{\mathrm{SS}}$ \emph{superselection sectors} $\qty{b^{\kappa}_{n}}=\qty[b^{\kappa}_{1}\dots b^{\kappa}_{N}]_{\kappa=1}^{\mathcal{N}_{\mathrm{SS}}}$, as in Eq.(3).
The explicit construction of candidates for initial states within the proposed dressed-site formalism requires that the left- and right-gauge irreps of each link share the same Casimir, as imposed in \cref{eq_link_symmetry}.
Dealing with staggered fermions \cite{Susskind1977LatticeFermions}, finite mass coupling values $m$ will favor configurations where even and odd sites host a complementary state of matter fields (vacuum and pair) respectively.
Therefore, good candidates for the initial state consist of staggered matter configurations where blocks of \(x\) empty (e) sites ($e_{1},\!\dots, e_{x}$) alternate with blocks of \(x\) doubly occupied (o) sites ($o_{1},\dots, o_{x}$).
These states have an initial matter imbalance, defined in Eq.(5), equal to 1, and their time evolution allows localization to be detected.

Practically, empty lattice sites are in a superposition of the states $\ket{1}$, $\ket{2}$, $\ket{7}$, and $\ket{8}$ of \cref{eq_extended_basis1,eq_extended_basis2} without matter fields.
Correspondingly, as for doubly occupied sites, we consider a superposition of the states $\ket{5}$, $\ket{6}$, $\ket{12}$, and $\ket{13}$ of \cref{eq_extended_basis1,eq_extended_basis2}.
According to the link symmetries in \cref{eq_link_symmetry}, we discard all superpositions in which neighboring site states do not share the same gauge irrep on the attached semi-links.
Any staggered initial (BG) state is then constructed with the following rules:
\begin{equation}
    \label{eq_superposition_rules}
    \begin{split}
        \ket{1}_{e_{1}}  & \otimes[\ket{1}_{e_{2}}+\ket{7}_{e_{2}}]\dots
        \ket{1}_{e_{x}}  \otimes[\ket{5}_{o_{1}}+\ket{12}_{o_{1}}]         \\
        \ket{2}_{e_{1}}  & \otimes[\ket{2}_{e_{2}}+\ket{8}_{e_{2}}]\dots
        \ket{2}_{e_{x}}  \otimes[\ket{6}_{o_{1}}+\ket{13}_{o_{1}}]         \\
        \ket{7}_{e_{1}}  & \otimes[\ket{2}_{e_{2}}+\ket{8}_{e_{2}}]\dots
        \ket{7}_{e_{x}}  \otimes[\ket{6}_{o_{1}}+\ket{13}_{o_{1}}]         \\
        \ket{8}_{e_{1}}  & \otimes[\ket{1}_{e_{2}}+\ket{7}_{e_{2}}]\dots
        \ket{8}_{e_{x}}  \otimes[\ket{5}_{o_{1}}+\ket{12}_{o_{1}}]         \\
        \\
        \ket{5}_{o_{1}}  & \otimes[\ket{5}_{o_{2}}+\ket{12}_{o_{2}}] \dots
        \ket{5}_{o_{x}}  \otimes[\ket{1}_{e_{1}}+\ket{7}_{e_{1}}]          \\
        \ket{6}_{o_{1}}  & \otimes[\ket{6}_{o_{2}}+\ket{13}_{o_{2}}] \dots
        \ket{6}_{o_{x}}  \otimes[\ket{2}_{e_{1}}+\ket{8}_{e_{1}}]          \\
        \ket{12}_{o_{1}} & \otimes[\ket{6}_{o_{2}}+\ket{13}_{o_{2}}] \dots
        \ket{12}_{o_{x}}  \otimes[\ket{2}_{e_{1}}+\ket{8}_{e_{1}}]         \\
        \ket{13}_{o_{1}} & \otimes[\ket{5}_{o_{2}}+\ket{12}_{o_{2}}] \dots
        \ket{13}_{o_{x}}  \otimes[\ket{1}_{e_{1}}+\ket{7}_{e_{1}}]
    \end{split}
\end{equation}
The simplest nontrivial staggered superposition of \emph{superselection sectors}  corresponds to $x=2$, where matter sites are in a sequence of two fully \emph{empty} ($e_{1}$ and $e_{2}$) and two fully \emph{occupied} sites ($o_{1}$ and $o_{2}$).
Such a configuration corresponds to the state defined in Eq.(3) and is used in all numerical simulations discussed in the main text.

Similar results can be obtained by choosing $x$ as any exact divisor of $N/2$, up to considering the half-chain domain wall configuration with empty sites ($e_{1} \dots e_{N/2}$) and the other one with fully occupied sites ($o_{N/2+1} \dots o_{N}$).
The main challenge of the latter state is its long relaxation timescale, which may strain numerical resources.

\subsection{MPS representation of the initial state}
In exact diagonalization, for an $N$-site lattice chain with periodic boundary conditions (PBC), any of these staggered initial states of the dynamics is made of a superposition of $4\cdot 2^{N-1}$ lattice configurations compatible with \cref{eq_extended_basis1,eq_extended_basis2}.
Correspondingly, the same superposition can be written as an exact matrix product state (MPS) of the form \cite{Fannes1992FinitelyCorrelatedStates, Klumper1991EquivalenceSolutionAnisotropic, Klumper1993MatrixProductGround}:
\begin{equation}
    \ket{\psi}=\sum_{i_1,..i_N} \mathrm{Tr} \left [ A^{(i_1)}_{1} A^{(i_2)}_{2}...A^{(i_N)}_{N} \right ] \ket{i_1,i_2,..i_N},
\end{equation}
where each matrix $A^{(i_{k})}_{k}$ has dimension $\chi_{k-1}{\times}\chi_{k}$ apart from the two edges, $\chi_{1}=1$ and $\chi_{N}=1$.
Under PBC, the superposition state of superselection sectors in Eq.(3) can be written as an MPS with bond dimension $\chi=4$ where the occupied sites take the form:
\begin{equation}
    A^{(i_{k})}_{\rm o}=\begin{pmatrix}
        \ket{5}      & \ket{12} & 0            & 0        \\
        \;\;\ket{13} & \ket{6}  & 0            & 0        \\
        0            & 0        & \ket{5}      & \ket{12} \\
        0            & 0        & \;\;\ket{13} & \ket{6}
    \end{pmatrix}\,,
\end{equation}
while for the empty sites
\begin{equation}
    A^{(i_{k})}_{\rm e}=\begin{pmatrix}
        \ket{1} & \ket{7} & 0       & 0       \\
        \ket{8} & \ket{2} & 0       & 0       \\
        0       & 0       & \ket{1} & \ket{7} \\
        0       & 0       & \ket{8} & \ket{2}
    \end{pmatrix}\,.
\end{equation}
The two edges ($\site=1,N$) read instead:
\begin{equation}
    A^{(i_{1(N)})}_{\rm o}=\begin{pmatrix}
        \ket{5}  \\
        \ket{12} \\
        \ket{13} \\
        \ket{6}
    \end{pmatrix} \quad\quad A^{(i_{1(N)})}_{\rm e}=\begin{pmatrix}
        \ket{1} \\
        \ket{8} \\
        \ket{7} \\
        \ket{2}
    \end{pmatrix}.
\end{equation}
This representation explains the initial value of the entanglement entropy in FIG.4 of the main text.

\section{Effective model at large couplings}
\begin{figure*}
    \includegraphics[width=0.9\textwidth]{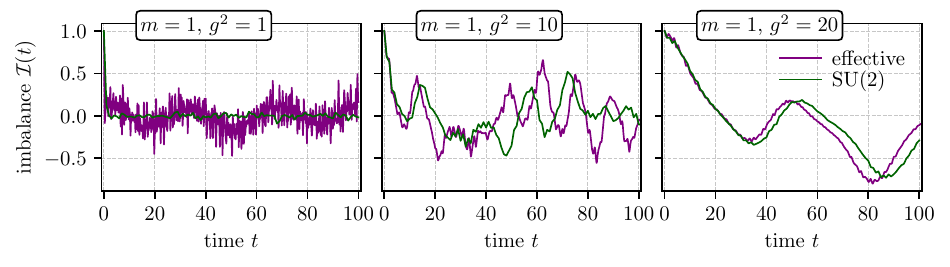}
    \caption{\textbf{Comparison with the effective model.}
        Imbalance as a function of time for different gauge couplings at fixed mass.
        The dynamics are initialized in the two-site state described by Eq.(4).
        The green lines refer to the truncated $\mathrm{SU}(2)$ model discussed in the main text, while the purple lines come from the effective model in \cref{eq_eff_model}.}
    \label{fig_effective_model}
\end{figure*}
In the strong-coupling limit, where $g^2\gg m,1$, the excitation of the links becomes energetically costly and can be adiabatically eliminated.
In the gauge-invariant sector with zero background charges, if the system is initialized in a pure matter state with no active gauge field, as in the state in Eq.(4), the model effectively reduces to the Hilbert subspace made of only $\ket{5}$ and $\ket{1}$ from \cref{eq_extended_basis1,eq_extended_basis2}.
Under these conditions, the model simplifies to a Heisenberg model with a staggered magnetic field along the z-axis:
\begin{equation}
    \ham_{\rm{eff}}=\frac{4}{g^2+m}\sum_n\hat{\vb*{\sigma}}_{n}\cdot\hat{\vb*{\sigma}}_{n+1}+m\sum_{\site}(-1)^{\site}\Sz_{\site}
    \label{eq_eff_model}
\end{equation}
where $\hat{\vb*{\sigma}}=(\Sx,\Sy,\Sz)$ are the Pauli matrices acting  on the states $|\uparrow\rangle=\ket{5}$ and $|\downarrow\rangle=|1\rangle$.

The accuracy of the effective model in capturing the system's dynamics can be observed in \cref{fig_effective_model}, where we show the time evolution of the imbalance in Eq.(5) at fixed mass $m=1$ for incremental gauge couplings $g^2$.
The agreement between Eq.(1) and \cref{eq_eff_model} becomes significant for $g^2\sim 20$, where the imbalance displays very long dynamics, even without a background-charge superposition.
Conversely, the effective model fails to describe the system evolution for smaller but moderately large couplings, $g^2\lesssim 10$.
In this regime, the system evolving in the zero-charge sector loses its initial-state configuration memory, and the occurrence of DFL induced by the superposition of charge sectors provides a nontrivial mechanism to escape thermalization.
In the large-mass regime $m\gg 1$, the dynamics in the zero charge sector in Eq.(4) cannot be reduced to just two states, making an effective spin model inapplicable.
Nevertheless, we discard this region of the couplings because the corresponding dynamics already fail to thermalize in the zero-background-charge sector.

\section{Thermal and long-time predictions}
\label{app_ensembles}
To clearly distinguish between the thermal and nonthermal behavior of the model under consideration, we compare the very long-time prediction with the one from the microcanonical ensemble.
For a unitary dynamics of a generic initial state expressed in the Hamiltonian eigenbasis, $\ket{\Psi(0)}=\sum_{s}C_{s}\ket{\Phi_{s}}$, the long-time behavior of a generic observable $\hat{O}$ corresponds to the expectation value predicted by the \emph{diagonal} ensemble (DE) \cite{Rigol2008ThermalizationItsMechanism}:
\begin{equation}
    \begin{split}
        \bar{O} & \equiv\lim_{T\to\infty} \frac{1}{T}\int_0^T dt \braket{\Psi(t)|\hat O|\Psi(t)}                                           \\
                & =\lim_{T\to\infty} \frac{1}{T}\int_0^T dt
        \sum_{s,p} C^{*}_{s}C_{p}\braket{\Phi_{s}|e^{it\hat{H}}\hat Oe^{-it\hat{H}}|\Phi_{p}}                                              \\
                & =\sum_{s,p} C^{*}_{s}C_{p}\braket{\Phi_{s}|\hat O|\Phi_{p}}\lim_{T\to\infty} \frac{1}{T}\int_0^T e^{i(E_{s}-E_{p})t} dt.
    \end{split}
    \label{eq_DE}
\end{equation}
If the Hamiltonian has no degeneracies, the oscillatory phases $e^{i(E_s-E_p)t}$ dephase for $s\neq p$, yielding:
\begin{equation}
    \bar{O}=\sum_{s}\abs{C_{s}}^2 \braket{\Phi_s|\hat O|\Phi_s}
    \equiv \Tr\{\hat O\, \hat{\varrho}_{\rm DE}\},
    \label{eq_nondegenerate_DE}
\end{equation}
where $\hat{\varrho}_{DE}=\sum_{s}\abs{C_s}^2 \ketbra{\Phi_s}$.
If instead the spectrum contains degenerate energy subspaces $\mathcal{H}_\alpha$ with projectors $\hat P_\alpha$, the long-time average retains all matrix elements within each degenerate block:
\begin{equation}
    \bar{O}=\sum_{\alpha}\Tr\left(\hat{P}_{\alpha}\, \ketbra{\Psi(0)}{\Psi(0)}\hat{P}_\alpha\right)\hat{O}.
    \label{eq_degenerate_DE}
\end{equation}
Correspondingly, the DE becomes block-diagonal
\begin{equation}
    \hat{\varrho}_{\rm DE}=\sum_{\alpha}\hat{P}_{\alpha}, \ketbra{\Psi(0)}{\Psi(0)}\hat{P}_{\alpha},
\end{equation}
and reduces to the usual diagonal form only when all energies are non-degenerate.

In any case, if the expectation value of the observable $\braket{\hat{O}(t)}$ approaches the long-time average $\braket{\hat{O}}_{DE}$ over long times, the system \emph{equilibrates}.
\begin{figure}
    \includegraphics[width=1\columnwidth]{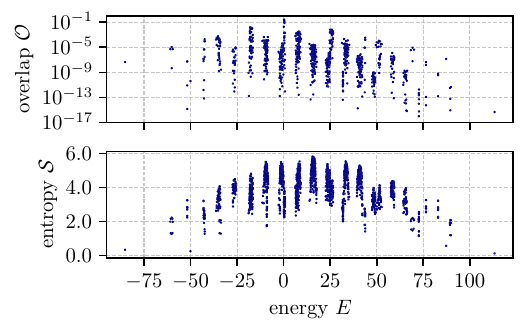}
    \caption{\textbf{Spectral properties of Fragmentation.}
        (left) Many-body spectrum analysis for Eq.(1) at $m\!=\!10$ and $g^{2}\!=\!2$: (top) the overlap $\mathcal{O}\!=\!\abs*{\braket{\Psi(0)|\Phi_s}}^2$ between the initial state in Eq.(4) and the spectrum eigenstates; (bottom) the corresponding half-chain bipartite entanglement entropy for each eigenstate of the spectrum.
        (right) Time evolution of the initial state in Eq.(4): (top) measure of the fidelity with the initial state and (bottom) half-chain bipartite entanglement entropy.}
    \label{fig_fragmentation_spectrum}
\end{figure}
Correspondingly, the expected thermal behavior can be computed with the microcanonical ensemble average (ME)
\begin{equation}
    \braket{\hat{O}}_{\rm{ME}}=\Tr\{\hat{O}\hat{\varrho}_{\rm{ME}} \}\,,
\end{equation}
where the microcanonical state is defined as a superposition within the energy shell $[E_\text{q}\!-\!\delta E, E_\text{q}\!+\!\delta E]$, with the initial-state energy $E_\text{q}\!=\!\bra{\Psi(0)}\hat{H}\ket{\Psi(0)}$, containing $N_{E,\delta E }$ eigenstates:
\begin{equation}
    \label{eq_microcanonical_state}
    \hat{\varrho}_{\rm{ME}}=\frac{1}{N_{E,\delta E }}
    \sum_{\substack{s;\\|E_s-E_\text{q}|<\delta E}}\ketbra{\Phi_s}{\Phi_s}\,.
\end{equation}
Whenever the ME and DE predictions of generic observables coincide, the system, quenched on the chosen initial state, \emph{thermalizes}.
Conversely, it exhibits nonthermal behavior due to Hilbert-space fragmentation, scarring dynamics, or other exotic phenomena.

In the superposition of superselection sectors in Eq.(2), DE and ME predictions of the imbalance in Eq.(5) (and any other observable) are obtained as classical averages of the DE and ME values of each background-charge configuration.
Namely, we have:
\begin{equation}
    \braket{\hat{O}}_{\rm{ME/DE}}^{\rm{BG}}=\frac{1}{\mathcal{N}_{ss}}\sum_{\kappa=1}^{\mathcal{N}_{ss}}
    \braket{\hat{O}}_{\rm{ME/DE}}^{\qty[b^{\kappa}_{1}\dots b^{\kappa}_{N}]}
    \label{eq_average_sector}
\end{equation}

\section{Hilbert space fragmentation}
To characterize the nonthermal yet delocalized region of the phase diagram detected in FIG.1 and FIG.3 as Hilbert space fragmentation (HSF), we explore the spectral and dynamical properties of the gauge-invariant sector of Eq.(1) in this specific regime of the couplings.
In particular, to exclude the occurrence of QMB scarring dynamics, for each eigenstate in the many-body spectrum $\{\ket{\Phi_s}\}$, we compute the overlap $\mathcal{O}\!=\!\abs*{\braket{\Psi(0)|\Phi_s}}^2$ with the initial state and the half-chain bipartite entanglement entropy $\mathcal{S}$ of each eigenstate.
As shown in \cref{fig_fragmentation_spectrum}, the Hamiltonian spectrum is well fragmented into tiny different vertical energy bands which do not communicate with each other (they are separated in energy by $\Delta E\!\sim\!8$).
Such a spectrum is clearly \emph{not} dense as the one observed in QMB scars (see FIG.3 of \cite{Calajo2025QuantumManybodyScarring}).

\begin{figure}
    \includegraphics[width=\columnwidth]{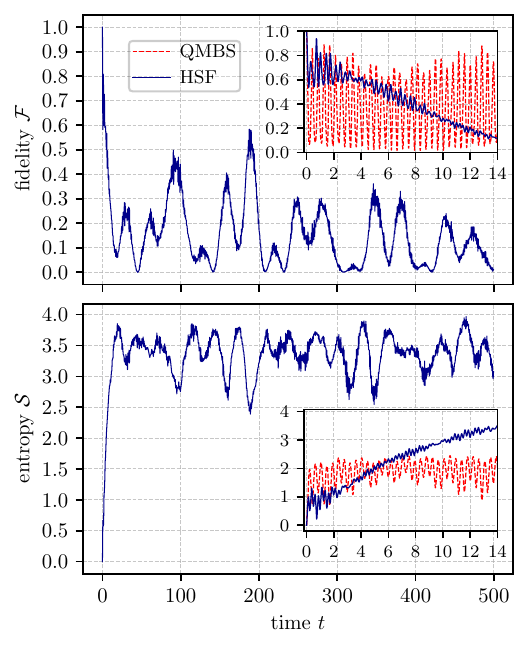}
    \caption{\textbf{Dynamical properties of Fragmentation.}
        Time evolution of the GI initial state in Eq.(4) at $m\!=\!10$ and $g^{2}\!=\!2$: (top) initial state fidelity $\mathcal{F}(t)\!=\!\abs{\braket{\Psi(t)|\Psi(0)}}^2$ as a function of time; (bottom) half-chain bipartite entanglement entropy scaling as a function of time.
        The inset panels compare the short-scale dynamical fluctuation of this fragmented regime with the persistent and periodic oscillations observed in QMB scars (red dashed lines) in \cite{Calajo2025QuantumManybodyScarring} at $m\!=\!1$ and $g^{2}\!=\!5$ when starting from the \emph{bare vacuum} state.}
    \label{fig_fragmentation_dynamics}
\end{figure}

A hallmark feature of QMB scars is that $\mathcal{F}(t)$ alternates between periodic (and persistent) revivals and complete memory loss of the initial state, where $\mathcal{F}(t)\simeq0$.
Such a feature is absent in the HSF case, whose dynamics are characterized by oscillations on two distinct timescales.
To visualize this different behavior, in \cref{fig_fragmentation_dynamics} we compare the return fidelity between the evolved state and the initial state, $\mathcal{F}(t)\!=\!\abs{\braket{\Psi(t)|\Psi(0)}}^2$ together with the time evolution of the corresponding half-chain bipartite entanglement entropy. We consider both the HSF case and the initial bare vacuum state studied in Ref.~\cite{Calajo2025QuantumManybodyScarring}, which exhibits scarred dynamics.
On a small timescale (see the inset panels of \cref{fig_fragmentation_dynamics}), comparable to the one of QMB scars, $\mathcal{F}(t)$ displays periodic \emph{damped} oscillations which, however, do \emph{not} erase memory of the initial state so that they cannot be truly called \emph{revivals} as for QMB scars.
Conversely, persistent, not periodic, revivals in the fidelity are observed on a timescale two orders of magnitude longer than the previous one, confirming the non-scarring origin of the observed dynamics.
The same small and large time-scales with irregular revivals of the fidelity are visible in the entanglement entropy, which rapidly saturates and clearly differs from the corresponding QMBS behavior (see \cite{Calajo2025QuantumManybodyScarring} and the lower inset panels of \cref{fig_fragmentation_dynamics}), where entropy oscillates with the same period of fidelity revivals while remaining bound to a much smaller value.

\begin{figure}
    \includegraphics[width=1\columnwidth]{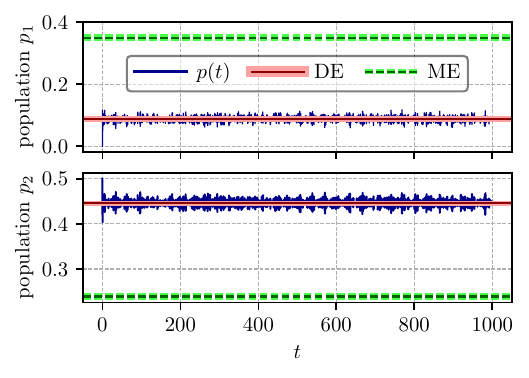}
    \caption{\textbf{Absence of thermalization.}
        Time evolution of the meson (top) and baryon (bottom) density populations $p_{1,2}(t)$ of Eq.(8) compared to the corresponding microcanonical (ME) and diagonal ensemble (DE) predictions at $g^{2}\!=\!2$, and $m\!=\!10$.}
    \label{fig_fragmentation2}
\end{figure}

To further characterize the lack of thermalization in this regime, we compare the real-time evolution of the meson and baryon densities in Eq.(8) with the corresponding predictions from ME and DE ensembles.
As shown in \cref{fig_fragmentation2}, the long-time dynamics of both populations predicted by the  DE do not converge to the thermal values predicted by the ME.
Therefore, despite this regime exhibiting localization with the SS initial state (see FIG.2), we do not consider it a genuine DFL, since it is inherently nonthermal (due to HSF) in the GI sector, even in the absence of a superposition of superselection sectors.

\section{Ergodic entanglement entropy scaling in the gauge invariant sector}
\label{sec_entanglement_GI}
\begin{figure*}
    \includegraphics[width=1\textwidth]{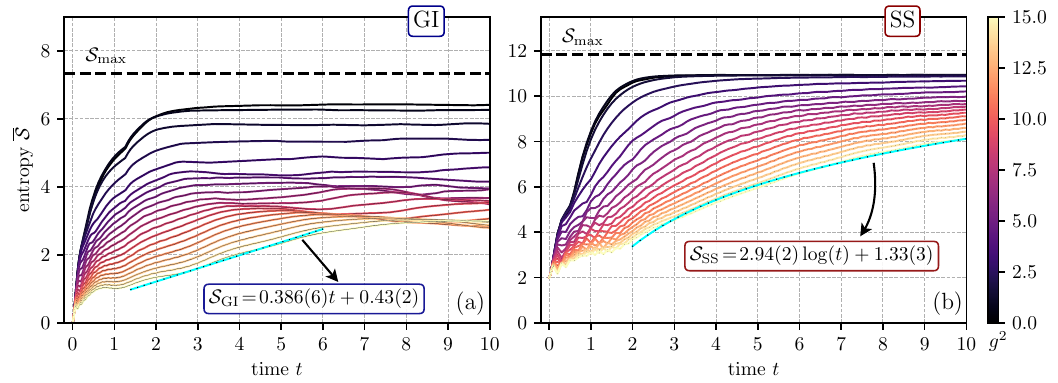}
    \caption{\textbf{Entanglement scalings in the GI and SS sectors.}
    Time-averaged half-chain bipartite entanglement entropy in base two as a function of time for fixed mass $m\!=\!1$, and a range of gauge couplings $g^{2}\!\in\![0, 15]$ in (a) the gauge-invariant (GI) and (b) superposition of superselection (SS) sectors.
    The cyan dashed lines correspond to (a) the linear fit of the ($g^{2}\!=\!15$) GI curve $\mathcal{S}_{\mathrm{GI}}=0.386(6)\cdot t + 0.43(2)$ and (b) the logarithmic fit of the ($g^{2}\!=\!15$) SS curve $\mathcal{S}_{\mathrm{SS}}=2.94(2)\log(t) + 1.33(3)$ respectively.
    The black dashed lines represent the maximal bipartite entanglement entropy value $\mathcal{S}_{\rm max}\!=\!\log {N_{\rm{subsys}}}$ reached when all the different $N_{\rm{subsys}}$ subsystem configurations (allowed in the GI or in the SS sector respectively) are equally probable.}
    \label{fig_entropy_GI}
\end{figure*}

In an ergodic regime, we expect quantum correlations to spread \emph{linearly} in time \cite{Kaufman2016QuantumThermalizationEntanglement}.
This is exactly what we observe for the GI sector across all coupling regimes compatible with ergodicity.
In \cref{fig_entropy_GI}, we display the bipartite entanglement entropy scaling within the GI sector as a function of time.
For a direct comparison, we show the GI entanglement's linear scaling, alongside the SS \emph{logarithmic} scaling already shown in FIG.4 of the main text.
To mitigate the effect of finite-size oscillations, similarly to what has been done for the imbalance in FIG.2, we consider the time-averaged entanglement entropy $\overline{\mathcal{S}}(t)=\frac{1}{\Delta t}\int^{t+\Delta t/2}_{t-\Delta t/2}d\tau\,\mathcal{S}(\tau)$, for an intermediate time window $\Delta t=2.5$.

As in the SS case discussed in FIG.4, large-$g^2$ couplings strongly penalize gauge excitations and suppress hopping processes.
The resulting constrained dynamics affects the spread of quantum correlations, yielding slower entanglement growth and lower saturation values than in the small-$g^{2}$ regime.
In addition to the thermal relaxation of local observables demonstrated in FIG.3, this confirms the ergodic nature of the GI sectors in the identified regimes of FIG.1.

\section{Exact diagonalization methods}
All numerical simulations presented in this work are performed using an Exact Diagonalization code \cite{Cataldi2026EdlgtExactDiagonalization, Cataldi2026DataCodeDisorderFree} that exploits the model's main symmetries.
Specifically, the Hamiltonian in Eq.(1) under PBC has the following symmetries: $\mathbb{Z}_{2}$-link symmetries (related to  \cref{eq_link_symmetry} and resulting from the previously discussed dressed-site formalism \cite{Cataldi2024Simulating2+1DSU2, Calajo2024DigitalQuantumSimulation, Calajo2025QuantumManybodyScarring}), which we resolve everywhere to satisfy SU(2) Gauss law; particle-number conservation, which we resolve by selecting the half-filling sector, corresponding to zero-baryon number density $\hat{N}_{b}\!=\!\hat{p}_{1}\!+\!2\hat{p}_{2}\!=\!0$; the momentum-symmetry, which we do not resolve as the initial SS and GI states in Eqs.(3) and (4) do not belong to a single sector; the spatial inversion symmetry, which maps each site $\site$ to its inverted partner $\site^{\prime}\!=\!(2\site[0]\!-\!\site)\!\!\mod\!N$ (site-centered) or $\site^{\prime}\!=\!(2\site[0]\!+\!1\!-\!\site)\!\!\mod \!N$.
If left unresolved, this invariance introduces degeneracy between positive-$k$ and negative-$k$ momentum states.
However, since our initial states belong to the $+1$ sector of the inversion parity, projecting the Hamiltonian onto this sector guarantees no degeneracy in the spectrum and allows us to compute the DE average using \cref{eq_nondegenerate_DE}.
Ultimately, thanks to the Gauss Law, we can independently access each superselection sector  $\qty{b^{\kappa}_{n}}=\qty[b^{\kappa}_{1}\dots b^{\kappa}_{N}]$ and recover the expectation values of the observables in Eq.(8) by averaging over the sectors as done in \cref{eq_average_sector}.
This is impossible when measuring the entanglement entropy discussed in FIG.4, since this quantity is not extensive across sectors.
A correct evaluation of the entropy requires simulating the initial state as a superposition over different superselection sectors, whose Hilbert space grows exponentially, $\sim 13^{N}$.
This prevents the system size from increasing and the finite-size scaling analysis from being performed via ED \cite{Cataldi2026EdlgtExactDiagonalization}.

As discussed earlier, one could alternatively encode the initial state as an MPS using tensor network methods.
Unfortunately, since our study aims to investigate ergodicity breaking phenomena, which require access to the thermal and very long time dynamics of the model where entanglement increases, MPS algorithms such as TEBD (Time-Evolving Block Decimation) \cite{Vidal2004EfficientSimulationOneDimensional} and TDVP (Time-Dependent Variational Principle) \cite{Haegeman2011TimeDependentVariationalPrinciple} are already inefficient at intermediate system sizes, such as $N=8$, where exact diagonalization methods can instead efficiently access the full energy spectrum to investigate the onset of fragmentation.
%\bibliography{bibliography}
\end{document}